\documentclass[aps,amssymb,amsmath,pra,twocolumn,superscriptaddress]{revtex4}
\usepackage{graphicx}
\usepackage{color}

\newcommand{\be}{\begin{equation}}
\newcommand{\ee}{\end{equation}}
\newcommand{\bc}{\begin{center}}
\newcommand{\ec}{\end{center}}
\newcommand{\bea}{\begin{eqnarray}}
\newcommand{\eea}{\end{eqnarray}}

\begin{document}
\title{Symmetry-noise interplay in quantum walk on an $n$-cycle}
\author{Subhashish Banerjee}
\email{subhashish@cmi.ac.in}
\affiliation{Raman Research Institute, Sadashiva Nagar, Bangalore
560 080, India}
\affiliation{Chennai Mathematical Institute, Padur PO, Siruseri 603103, India}
\author{R. \surname{Srikanth}}
\affiliation{Poornaprajna Institute of Scientific Research,
Devanahalli, Bangalore 562 110, India}
\affiliation{Raman Research Institute, Sadashiva Nagar, Bangalore, India}
\author{C. M. \surname{Chandrashekar}}
\affiliation{Institute for Quantum Computing, University of Waterloo
ON N2L 3G1, Canada}
\affiliation{Perimeter Institute for Theoretical Physics, Waterloo, ON, 
N2J 2W9,Canada}
\author{Pranaw \surname{Rungta}}
\affiliation{CHEP, Indian Institute of Science, Bangalore- 560 012, India.}
\begin{abstract}

Augmenting the  unitary transformation which generates  a quantum walk
by  a generalized  phase gate  $G$ is  a symmetry  for both  noisy and
noiseless quantum  walk on  a line,  in the sense  that it  leaves the
position  probability distribution  invariant. However,  this symmetry
breaks down in  the case of a quantum walk on  an $n$-cycle, and hence
can be regarded as a probe of the walk topology.  Noise, modelled here
as  phase flip and  generalized amplitude  damping channels,  tends to
restore  the symmetry  because  it classicalizes  the walk.   However,
symmetry restoration  happens even in  the regime where the  walker is
not  entirely  classical,  because   noise  also  has  the  effect  of
desensitizing  the operation  $G$ to  the walk  topology.   We discuss
methods  for  physical  implementation,   and  talk  about  the  wider
implications to condensed matter systems.

\end{abstract}

\maketitle
\preprint{Version}
\section{Introduction}
\label{intro}

Classical  random  walks  (CRWs)  has found  broad  applications---from
randomized  algorithms  to   the  understanding  of  condensed  matter
systems~\cite{Barber}.  Unlike CRW, quantum walks (QWs)~\cite{feynman}
involve   a   superposition   of   states,  being   unitary,   thereby
simultaneously exploring multiple possible  paths of a walker, and the
amplitudes corresponding to different  paths are then interfered via a
measurement to arrive at a probabilistic result.  This makes QW spread
quadratically  faster than CRW.   The quadratic  advantage of  QW, for
instance,  is  exploited to  speedup  the  spatial  search variant  of
Grover's search  algorithm~\cite{Ambainis}.  A single-particle quantum
lattice  gas automata (QLGA)  can also  be shown  to be  equivalent to
QW~\cite{meyer}.  Experimental implementation  of QW has been reported
\cite{ryan},  and various  other schemes  have been  proposed  for its
physical   realization~\cite{travaglione,    chandra06},   which   has
motivated us  to study QW in an  open quantum system~\cite{qrw1,wl73}:
since environmental  effects (noise) will necessarily  tend to destroy
the coherent superposition of  states central to QW, thus transforming
it to CRW.

This paper studies  QW on  an  $n$-cycle, which  is
subjected   to  the   following  noise   processes:  the   phase  flip
(decoherence   without   dissipation)~\cite{bg07},   and   generalized
amplitude damping  channels (decoherence with dissipation)~\cite{qrw1,
nc00,br06,wineland},  which are  of  relevance to  studies in  quantum
optics and condensed matter systems.   The latter type of noise in the
QW   context  has been  studied   by  us   in
Ref. \cite{qrw1}. We study the transition of  a noisy QW on
an $n$-cycle to  a CRW in a different way: we  identify a certain symmetry
operation (defined below)  that is sensitive to the  walk topology, in
the sense that the symmetry holds for QW on a line but not for that on
a cycle.   The difference arises due to  the fact that unlike  QW in a
line, the  walk on an $n$-cycle involves  interference between forward
and  backward  propagating   wavefunctions.  Noise  tends  to  restore
symmetry both  by classicalizing the  walk and also  desensitizing the
symmetry operation as a topology probe for the QW.

This paper is organized as follows. In Sec.  \ref{qwpg}, we define the
discrete time  QW on  an $n$-cycle and  introduce a  generalized phase
gate which is a symmetry  operation and show the breakdown in symmetry
when the  QW is implemented on  an $n$-cycle. In  Sec.  \ref{nqw}, the
effect of  noise on  the QW  and its influence  on the  restoration of
symmetry  is discussed.   The implication  of this  work  to condensed
matter systems is discussed  in Sec.  \ref{cms} before conclusions are
made in Sec. \ref{conc}.

\section{Quantum walk on an $n$-cycle with generalized phase gate}
\label{qwpg}
Consider a particle (a  qubit) which is executing a discrete
time  QW in  one dimension,  and its  internal states  $|0\rangle$ and
$|1\rangle$ span ${\cal  H}_c$, which is referred as  the coin Hilbert
space. The  allowed position states  of the particle  are 
$|x\rangle_p$,
which  spans  ${\cal  H}_x$,  where  $x  \in  {\bf  I}$,  the  set  of
integers (the subscript $p$ in the ket is used to distinguish the position
kets from the internal states, which are represented subscriptless). 
In an  $n$-cycle walk, there are $n$  allowed positions, and
in  addition  the  periodic  boundary condition  
$|x\rangle_p=|x ~{\rm
mod}\;n\rangle_p$  is imposed.   A $t$  step coined  QW is  generated by
iteratively applying a unitary operation $W$ which acts on the Hilbert
space ${\cal H}_c\otimes {\cal H}_x$:
\begin{equation}
\label{eq:walk}
|\psi_t\rangle=W^t|\psi_0\rangle\;,
\end{equation}
where  $|\psi_0\rangle      =
(\cos(\theta_0/2)|0\rangle+ \sin(\theta_0/2) e^{i\phi_0}
|1\rangle)|0\rangle_p$ is an arbitrary initial
state of the particle and $W\equiv U\ B(\xi,\theta,\zeta)$.
The $B(\xi,\theta,\zeta)$ is an arbitrary $SU(2)$ coin toss operation which acts on the
coin space given by 
\be
\label{eq:su2}
B(\xi,\theta,\zeta)=\left(
\begin{array}{clcr} e^{i\xi}\cos(\theta) & & e^{i\zeta}\sin(\theta) \\
e^{-i\zeta}   \sin(\theta)  &  &   -e^{-i\xi}\cos(\theta)  \end{array}
\right).
\ee
The matrix  $B(\xi,\theta,\zeta)$, whose elements are written
as $B_{jk}$, controls the  evolution of the walk, with the Hadamard
walk   corresponding   to   $B(0^\circ,45^\circ,0^\circ)$.   The  $U$   is   a   unitary
controlled-shift operation:
\be
\label{eq:cs}
U\equiv   |0\rangle\langle0|\otimes\sum_x|x-1\rangle_p
\langle  x|_p   +
|1\rangle\langle 1|\otimes\sum_x|x+1\rangle_p\langle x|_p.
\ee
The probability  to find the particle  at site $x$ after  $t$ steps is
given      by
\be
 \label{eq:prpb}
p(x,t)       =      \langle      x|_p{\rm      tr}_c
(|\psi_t\rangle\langle\psi_t|)|x\rangle_p.
\ee
 Now,  given an element from
the  $2$-parameter   group 
\be
\label{eq:pg}
G(\alpha,\beta)=  \left(\begin{array}{ll}
e^{i\alpha} & 0 \\ 0 & e^{i\beta}\end{array}\right),
\ee
which represents
a generalized phase gate acting on  the ${\cal H}_c$, we find that the
operation $W  \longrightarrow GW$ leaves  the probability distribution
$p(x,t)$  of the particle  on the  line invariant;  hence the  walk is
symmetric  under the  operation 
\be
\label{eq:pgope}
G(\alpha,\beta):  |j\rangle \mapsto
e^{i(\overline{j}\alpha +  j\beta) }|j\rangle
\ee
 for  $|j\rangle$ in the
computational basis (eigenstates of the Pauli operator $\sigma_z$) and
$j=0,1$.  The physical significance of $G$ is that it helps identify a
family  of QWs  that are  equivalent  from the  viewpoint of  physical
implementation,  which  can sometimes  allow  a significant  practical
simplification \cite{qrw1}.   For example, suppose  the application of
the  conditional  shift is  accompanied  by  a  phase gate.  The  walk
symmetry implies  that this  gate need not  be corrected  for, thereby
resulting in  a saving of  experimental resources. The inclusion  of a
phase gate on the coin operator  is equivalent to a phase gate at each
lattice site  in the  sense of  QLGA, with the  physical meaning  of a
constant potential. The
evolution rules for single-particle  QLGA can be classified into gauge
equivalent  classes, there  being a  difference between  the  class of
rules for periodic  ($n$-cycle) and non-periodic 1-dimensional lattice
and this  feature can  be exploited to  distinguish between  these two
spatial topologies \cite{meyer1}.

\begin{figure}
\includegraphics[width=8.2cm]{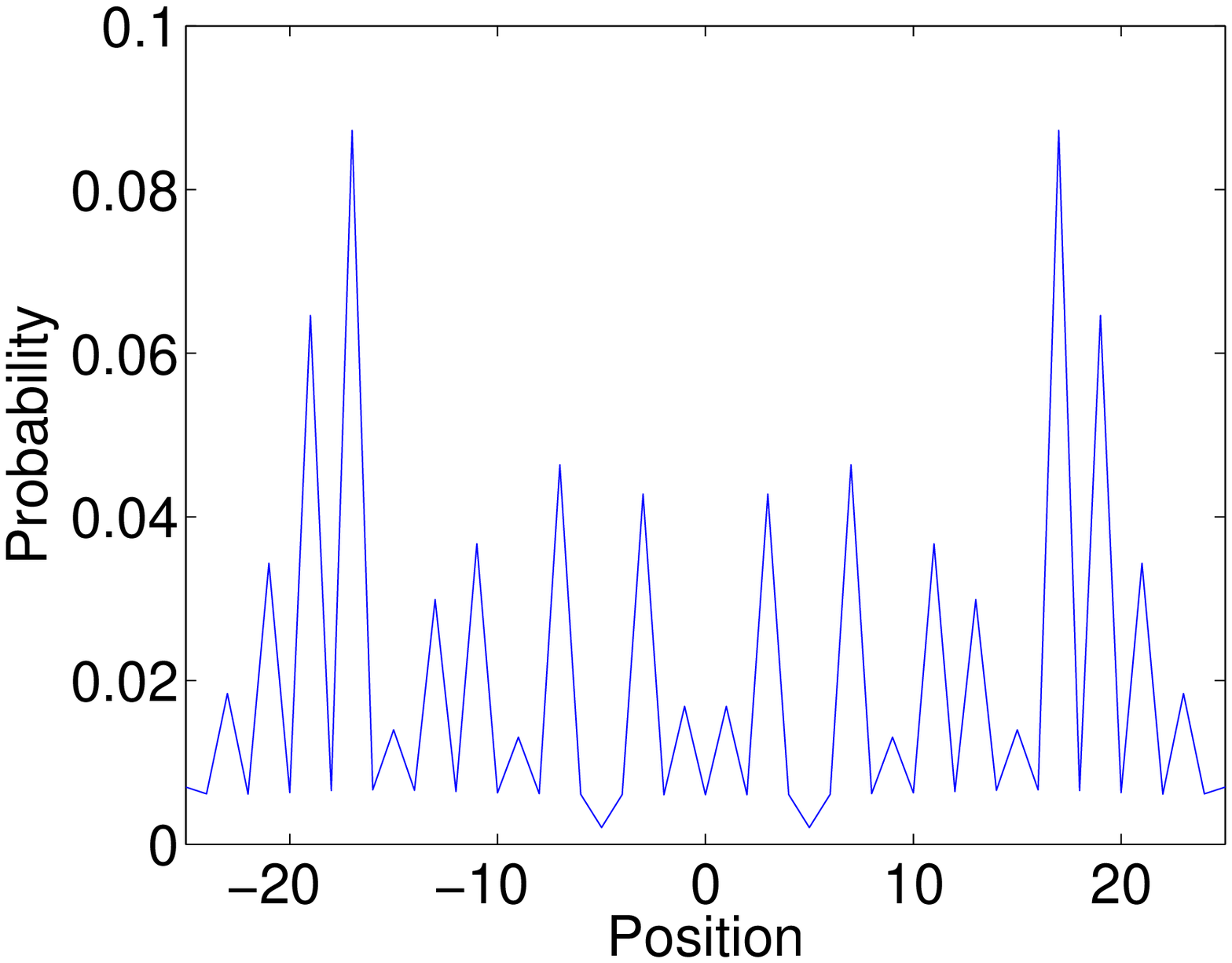}\\
$~~~$(a)\\
\includegraphics[width=8.2cm]{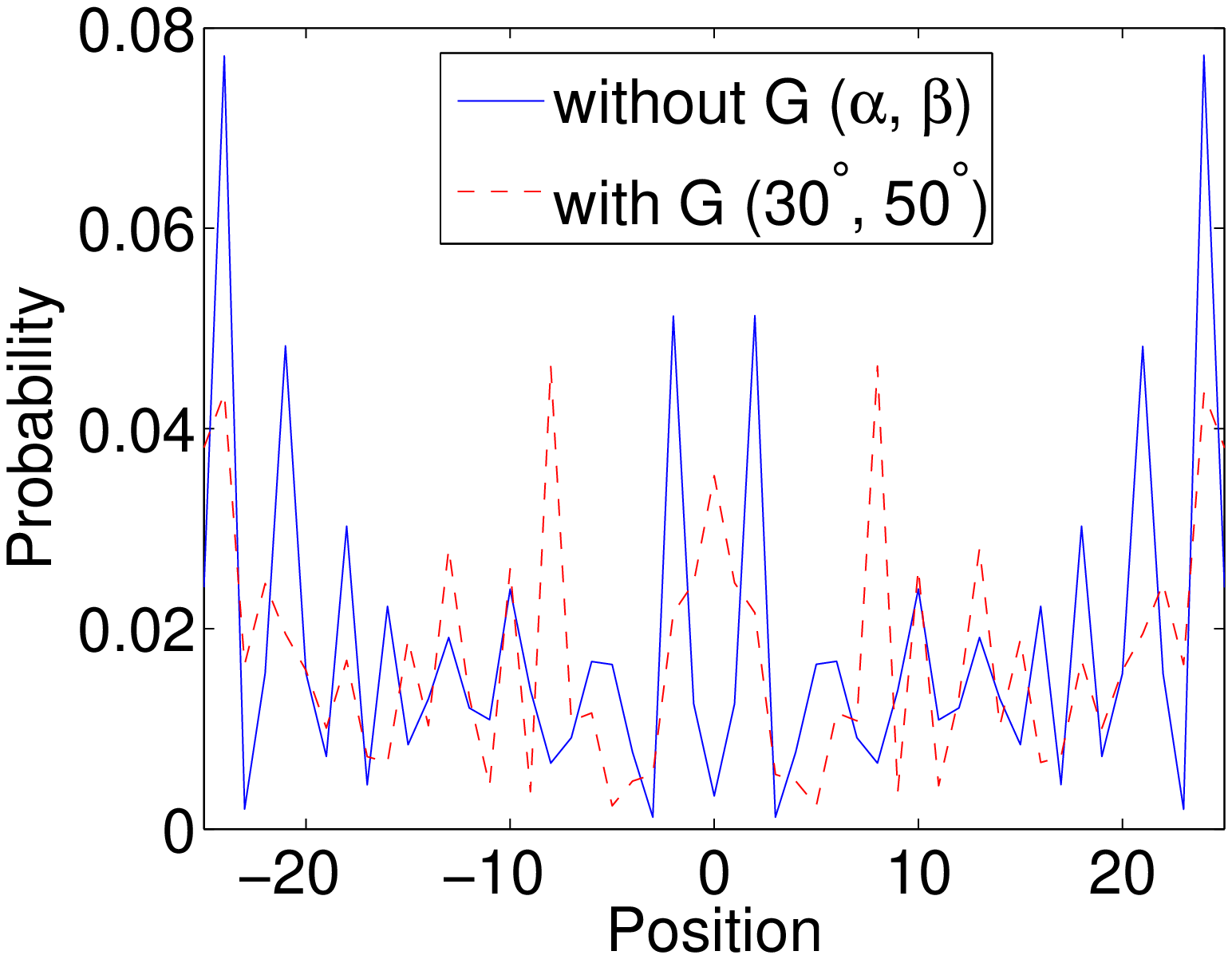}\\
$~~~$(b)
\caption{(Color  online)  Position   probability  distribution  for  a
  Hadamard  walk, $B(0^{\circ},  45^\circ,0^{\circ})$ on  (a) a line,
and (b) an $n$-cycle, with
  ($n=51$)  and  initial    state    $(1/\sqrt{2})(|0\rangle   +
|1\rangle)$, for the unitary   case    ($\gamma_0=0$). Each figure
presents the distribution with and without being subjected to
the phase  operation $G(30^\circ,50^\circ)$.
In (a), there is perfect symmetry, since both distributions coincide.
In (b) the two  plots  do  not   overlap,  indicating the  breakdown  of  the
  symmetry.}
\label{fig:noisy_amp}
\end{figure}

It turns  out that in  the case of  QW on an $n$-cycle,  this symmetry
breaks down.   To see this, we  note that the  $t$-fold application of
the operation $GUB$ on  a particle with initial state $|\psi_0\rangle$
on the line and on an $n-$cycle produces, respectively, the states
\begin{widetext}
\begin{subequations}
\label{eq:gsup}
\begin{eqnarray}
(GUB)^t|\psi_0\rangle =
\sum_{j_1,j_2,\cdots,j_t}e^{i(\overline{J}_t\alpha + J_t\beta)} 
B_{j_t,j_{t-1}}\cdots B_{j_2,j_1}
(B_{j_1 0}a + B_{j_1 1}b)|j_t\rangle|2J_t-t\rangle_p, \label{eq:gsupa} \\
(GUB)^t|\psi_0\rangle = 
\sum_{j_1,j_2,\cdots,j_t}e^{i(\overline{J}_t\alpha + J_t\beta)}  
B_{j_t,j_{t-1}}\cdots B_{j_2,j_1} 
(B_{j_1 0}a + B_{j_1 1}b)|j_t\rangle|2J_t-t\mod n\rangle_p, \label{eq:gsupb}
\end{eqnarray}
\end{subequations}
\end{widetext}
where  $J_t =  j_1+j_2+\cdots+j_t$ and  $\overline{J}_t$ is  a bitwise
complement  of  $J_t$.  All  terms  in superposition  (\ref{eq:gsupa})
contributing  to the  probability  to  detect the  walker  at a  given
position   $x  =   2J_t-t$   have  the   {\it   same}  phase   factor,
$e^{i(\overline{J}_t\alpha  + J_t\beta)}$,  which is  fixed by  $J_t =
(x+t)/2$ (where,  it may be noted, $x$  and $t$ are both  even or both
odd).  Thus, this factor does not affect the probability to detect the
walker at $x$, whence the symmetry.  In the case of QW on an $n$-cycle
the    breakdown    of     the    symmetry~\cite{qrw1},    see    Fig.
\ref{fig:noisy_amp}, can  be attributed to the topology  of the cycle,
which introduces a periodicity in the walker position (determined by a
congruence relation  with modulus given  by the number of  sites), but
not in the phase of the  superposition terms.  As a result, fixing $x$
fixes $J_t \mod n = (x+t)/2 \mod n$, but not $J_t$ itself, so that the
phase terms  in the superposition  Eq. (\ref{eq:gsupb}) do  not factor
out globally. Thus if $\alpha$ or $\beta$ is non-vanishing and $\alpha
\ne \beta$, then  in general the symmetry $G$ is  absent in the cyclic
case.

We  quantify the  breakdown in  symmetry  by means  of the  Kolmogorov
distance  (or,   trace  distance   \cite{nc00}),  given  by 
\be
\label{eq:kd}
d(t)  =
(1/2)\sum_x    |p(x,t)-q(x,t)|,
\ee
between    the    walker   position
distributions obtained without and  with the symmetry operation, given
by $p(x,t)$ and $q(x,t)$, respectively.  The breakdown in symmetry for
a   noiseless  cyclic   QW  is   depicted   by  the   bold  curve   in
Fig.  \ref{fig:kd_uni} as  a function  of the  number of  turns $\tau$
(where $t=\tau s$, with $n =2s+1$).

\section{Effect of noise and symmetry restoration}
\label{nqw}

We now  describe the $n$-cycle quantum walk  on the particle, a  two level
system, when subjected  to  noise. The situation is modeled as  an  interaction with  a
thermal  bath, characterized  by phase  damping \cite{nc00,qrw1}  or a
generalized  amplitude  damping  channel,  the  latter  process  being
represented by the following Kraus operators \cite{br06}:
\begin{eqnarray}
\label{eq:gbmakraus}
\begin{array}{ll}
E_0 \equiv \sqrt{\kappa}\left[\begin{array}{ll}
\sqrt{1-\lambda(\Delta)} & 0 \\ 0 & 1
\end{array}\right]; &
E_1 \equiv \sqrt{\kappa}\left[\begin{array}{ll} 0 & 0 \\ \sqrt{\lambda(\Delta)} & 0
\end{array}\right];  \\
E_2 \equiv \sqrt{1-\kappa}\left[\begin{array}{ll} 1 & 0 \\ 0 &
\sqrt{1-\lambda(\Delta)}
\end{array}\right];  &
E_3 \equiv \sqrt{\frac{1-\kappa}{\kappa}}E_1^\dag,
\nonumber \\
\end{array}
\end{eqnarray}
where $0 \le \kappa \le 1$, $\lambda(\Delta) = 1-e^{-\gamma_0(2 N_{\rm
th}+1)\Delta}$,  and $\kappa  \equiv  \frac{N_{\rm th}+1}{2N_{\rm  th}
+1}$, with  $N_{\rm th} \equiv  (\exp(\hbar\omega/k_B T)-1)^{-1}$, $T$
is temperature, $\gamma_0$ is a measure of the strength of coupling to
the  environment,   and  $\Delta$  is  the  duration   for  which  the
environment  is  modeled  to  interact  with the  coin.   The  density
operator  $\rho_c$  of the  coin  evolves  according to  \mbox{$\rho_c
\rightarrow \sum_j  E_j \rho_c E_j^\dag$}.  The full  evolution of the
walker, described  by density operator $\rho(t)$, is  given by $\sum_j
E_j(W \rho(t-1) W^\dag)E_j^\dag$, where  the $E_j$'s are understood to
act only in the coin space.

\begin{figure}
\includegraphics[width=8.0cm]{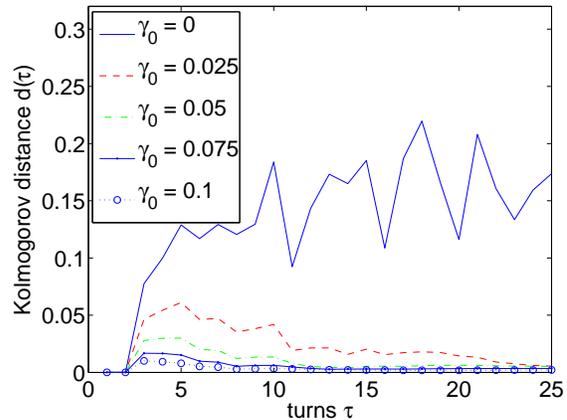}
\caption{(Color  online)  Kolmogorov  distance $d(\tau)$  against  the
  number of turns ($\tau$) of the cyclic QW in the noiseless and noisy
  case with $n=51$.  For the unitary case ($\lambda=0$; bold line) the
  walk  becomes increasingly  asymmetric  as the  number  of turns  is
  increased, until about 7--10 turns, after which it fluctuates around
  $d\approx0.15$.  The  plots represent generalized  amplitude damping
  noise at different  noise levels at temperature $T  = 3.5$ (in units
  where   $\hbar    \equiv   k_B   \equiv    1$),   $\Delta=0.1$   and
  $\theta=30^{\circ}$.  The  walker is evolved with  the initial state
  parameters (in degrees) $\theta_0=30^\circ$, $\phi_0=40^\circ$, with
  $B(20^\circ, 10^\circ,30^\circ)$ and $G(40^\circ, 50^\circ)$.}
\label{fig:kd_uni}
\end{figure}

The curves  in Fig. \ref{fig:kd_uni}  plot $d(\tau)$ as a  function of
turns  in  the   case  of  unitary  and  noisy   QW  (parametrized  by
$\gamma_0$), and demonstrate the  gradual restoration of symmetry with
time on account of the noise.  Although the figure employs generalized
amplitude damping  noise, qualitatively the same behavior  can be seen
for a phase damping noise.   Here, a general feature is that $d(\tau)$
is non-zero  when $\tau<2$, being then  equivalent to (noisy)  QW on a
line.  Thereafter, $d(\tau)$ at first increases with increasing turns,
being dominated by unitary evolution, and eventually falls down, being
dominated by  noise.  It is  observed that for sufficiently  low noise
levels,  the time  at which  this  turnover in  slope happens  remains
constant,   for   given   $\theta$.    This  is   depicted   in   Fig.
\ref{fig:kd_uni}  for  the case  of  a  generalized amplitude  damping
channel corresponding  to a fixed temperature  and varying $\gamma_0$.
However, we  note that for  strong enough noise, the  turnover happens
earlier.
\begin{figure}
\includegraphics[width=8.4cm]{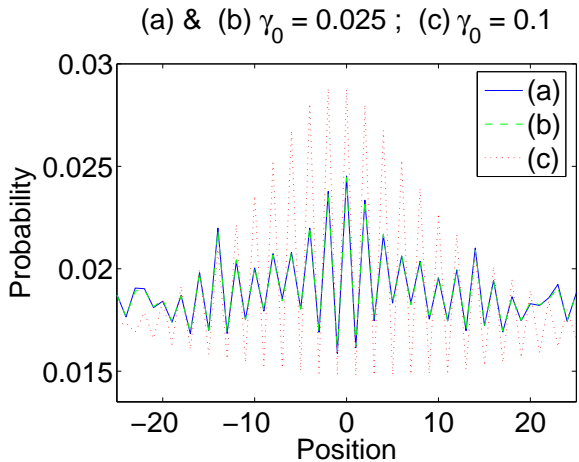}
\caption{(Color  online)  Position   probability  distribution  for  a
  Hadamard  walk  $B(0, 45^\circ,0)$  on  an  $n$-cycle ($n=51$)  with
  initial   state  $(1/\sqrt{2})(|0\rangle   +   |1\rangle)$  when
  subjected to  generalized amplitude damping  noise with $\Delta=0.1$
  and finite $T$  ($= 6.0$) for $\tau = 11$.  (a)  and (b) $\gamma_0 =
  0.025$, (b)  is the distribution  for the QW augmented  by operation
  $G(30^\circ,50^\circ)$;  note  the  walk  remains quantum  yet  with
  symmetry  almost  completely  restored;  (c)  classicalized  pattern
  (indicated by the regular envelope) obtained with larger noise level
  corresponding to $\gamma_0=0.1$.}
\label{fig:sym_restore}
\end{figure}

Typical    noisy   probability    distributions   are    depicted   in
Fig. \ref{fig:sym_restore}(a),(b) at an instant where the symmetry has
been  almost fully  restored even  when the  walk is  well  within the
quantum    regime.   Fig.   \ref{fig:sym_restore}(c)    represents   a
classicalized  distribution, indicated by  the regular  envelope (that
will eventually turn into a uniform distribution).

We define coherence ${\bf C}$ as  the sum of the off-diagonal terms of
states  in ${\cal  H}_c \otimes  {\cal H}_p$,  where ${\cal  H}_c$ and
${\cal  H}_p$ are  the  internal  and position  Hilbert  space of  the
quantum walker, respectively.   If the state of the  quantum walker is
$\rho  =   \sum_{ab;jk}\alpha_{ab;jk}|a\rangle|j\rangle_p
\langle  b|\langle k|_p$,
where $|a\rangle$,  $|b\rangle \in {\cal  H}_c$ and  
$|j\rangle_p$, $|k\rangle_p \in  {\cal H}_p$, then  
\be
{\bf C}  \equiv \left(\sum_{a  \ne b,j  =k} +
\sum_{a,b,j \ne  k} + \sum_{a \ne  b,j \ne k}\right)|\alpha_{ab;jk}|,
\ee
the sum of the absolute values  of all off-diagonal terms of $\rho$ in
the computational-position  basis.  The coherence  function is defined
as the  quantity $C(m)$, where  $m \in \{1,2,\cdots,M\}$,  obtained by
partitioning ${\bf C}$ into $M$ intervals of size $s/M$, such that for
the  $m$th  interval $(m-1)(s/M)  \le  |j-k|  < m(s/M)$.   Physically,
$C(m)$ is a measure of coherence  between two points on a (in general,
noisy) quantum walker, as a  function of their mutual separation.  Let
$C_0(m)$  represent  the   coherence  function  of  the  corresponding
noiseless walk. At any turn $\tau$, we define the normalized coherence
function  by $c(m) \equiv  C(m)/C_0(m)$, and,  analogously, normalized
Kolmogorov distance by $D(\tau) \equiv d(\tau)/d_0(\tau)$.

\begin{figure}
\includegraphics[width=8.0cm]{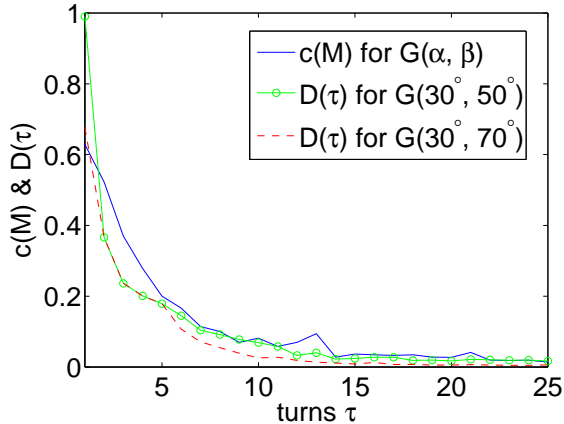}
\caption{(Color  online) 
(Color  online) Normalized  coherence function  $c(M)$ (bold)
   and  the normalized  Kolmogorov  distance $D(\tau)$  (dashed) for  a
   cyclic QW  as a  function of turns  $\tau$. An overall  reduction of
   both  $c(M)$ and  $D(\tau)$ with  time is  seen for  the generalized
   amplitude  damping noise  characterized by  temperature  $T=3.5$ and
   $\Delta=0.1$. The  Hadamard walk is evolved with  the initial state
   parameters (in degrees) $\theta_0=45^\circ$, $\phi_0=0^{\circ}$, and
   $G(\alpha, \beta)$. The  line-with-circle plot  represents $D(\tau)$ with  $G(30^\circ,
50^\circ)$ and dashed line represents for $G(30^\circ,  70^\circ)$.
The  $c(M)$  (solid line) remains  roughly the same for both pairs of
$\alpha$ and $\beta$.  For a clear depiction of the notion of symmetry 
restoration in the
walk even in the quantum regime, cf. Figure \ref{fig:sym_restore}.}
\label{fig:coherence}
\end{figure}

Since  noise tends  to destroy  superpositions, and  the  breakdown in
symmetry is  essentially a phenomenon of superposition  of the forward
and  backward waves,  noise tends  to restore  symmetry, as  seen from
Fig.   \ref{fig:sym_restore}.    This   is   brought   out   by   Fig.
\ref{fig:coherence} for two possible values of $G$.
In the figure, in spite of its considerable spikiness, the
bold  curve,  representing $c(m=M)$,  shows  an  overall fall.
A similar trend as depicted in this figure, has been numerically 
checked for various other values of $G$. This
raises the question whether symmetry  restoration of the cyclic QW can
be considered as a good indicator of classicalization.  
Here we note that from Fig. \ref{fig:sym_restore}(a),(b) the
probability distribution  pattern is seen to be  clearly quantum, even
though symmetry has been almost fully restored.
This is suggestive of the notion that
that symmetry tends to be restored {\em even} in
the regime  where the walk  still possesses some quantum  features.

The reason $D(t)$  is not a faithful indicator  of classicalization of
the walk has to do with the  effect of noise on the sensitivity of the
symmetry  operation $G(\alpha,\beta)$  to  the topology  of the  path.
Since  this operation  senses  the  closure of  the  path through  the
superposition of  the forward and  backward waves, the  suppression of
superposition through noise will also have the effect of desensitizing
the operation  to the  closure of the  path, thereby moving  the noisy
cyclic QW  towards a noisy QW on  a line from the  perspective of this
operation, before further classicalization transforms it into a cyclic
CRW.  And as shown in  Ref.  \cite{qrw1}, all the above symmetries are
respected by  a (noisy) quantum  walk on a  line, both in the  case of
phase damping noise, which in NMR nomenclature \cite{bram} is called a
$T_2$ process,  and generalized amplitude  damping, which is  a $T_1$,
$T_2$ process \cite{bram}. This  brings out the point that decoherence
($T_2$  process)  is  the  principal  mechanism  responsible  for  the
restoration of  symmetries.  It also highlights  the interplay between
topology  and noise  in a  quantum walk  on an  $n-$cycle.   A similar
interplay may be expected also in the case of QW with other nontrivial
topologies.

The experimental study of the  decoherence and decay of quantum states
of  a trapped  atomic ion's  harmonic motion  subjected  to engineered
reservoirs, both of the phase damping and amplitude damping kind, have
been reported in \cite{wineland}.  The phase reservoir is simulated by
random variation  of the trap frequency $\omega$  without changing its
energy (non-dissipative),  while the amplitude  reservoir is simulated
by   random   electric   field    along   the   axis   of   the   trap
(dissipative).     Coupling     the     reservoirs     reported     in
Ref.  \cite{wineland}  to the  scheme  presenting  the combination  of
pulses required to implement  a QW on a line and on  a cycle in an ion
trap \cite{travaglione}  provides a  convenient set up  to demonstrate
the symmetry-noise interplay.

The interplay  between geometry and decoherence has  been noted before
in  the  case  of  delocalized  bath modes  \cite{wei98},  as  against
localized bath modes \cite{wei98,all95}.   This is of relevance as the
noise processes considered here  \cite{br06,qrw1} are described by the
interaction of the system with delocalized bath modes.

\section{Implications to condensed matter systems}
\label{cms}

Breaking of  the symmetry  due to the  change in walk  topology causes
long-range correlations to develop, in analogy to the hydrodynamics of
ordered   systems   such   as   spin   waves   in:   ferromagnets,  
antiferromagnets   (where   it  is   the   spin   wave  of   staggered
magnetization),  second  sound in  He$^{3}$,  nematic liquid  crystals
\cite{hydro}.    Here  the   correlations  may   be   identified  with
symmetry-broken  terms  (whose   measurement  probability  depends  on
$\alpha$ or $\beta$ in the walk augmented by $G(\alpha,\beta)$) in the
superposition of  the quantum walker. One finds  that correlations are
set up  rapidly over large distances  with increase in  the winding of
the walker, until symmetry is broken throughout the cycle. However, as
noted  above,   the  randomization   produced  by  noise   causes  the
reappearance of  symmetries.  The  symmetry breaking and  the symmetry
restoring  agents  are  thus   different,  the  former  given  by  the
topological transition from  a line to an $n$-cycle,  the latter being
the noise-induced randomization.

Coherence is also widely used to understand quantum phase transitions,
the transition from  superfluid to Mott insulator state  in an optical
lattice   being   one  specific   example   \cite{jaksch}.   In   Ref.
\cite{qwqpt}  the   quantum  phase  transition  using  QW   in  a  one
dimensional optical lattice has been discussed.  Using various lattice
techniques,  desired geometries to  trap and  manipulate atoms  can be
created.   In   most  physical   situations  one  deals   with  closed
geometries. The  characteristics of the $n-$cycle  walk, in particular
the   re-appearance   of   the   symmetry  (implying   a   family   of
implementationally  equivalent noisy  cyclic QWs)  while still  in the
quantum regime,  presented here could  be of direct relevance  to such
situations.   

The ubiquity of the ideas developed in this paper can be seen from the
fact that the quantum dynamics of a particle on a ring (cycle) subject
to decoherence along  with dissipation finds its place  in the physics
of quantum dots.  The effective action of a quantum dot accounting for
the  joint   effect  of  charging  and  coupling   to  an  environment
\cite{as82} mirrors the behavior of the quantum dynamics of a particle
on  a  ring  (cycle)   subject  to  a  dissipative  damping  mechanism
describing  the dissipation of  the energy  stored in  dynamic voltage
fluctuations  into   the  microscopic   degrees  of  freedom   of  the
quasi-particle continuum.   In the absence of  dissipation, the action
describes the  ballistic motion of a  quantum particle on  a ring. The
ring topology  reflects the  $2\pi$-periodicity of the  quantum phase,
which  is in  turn  related  to the  quantization  of charge,  thereby
highlighting  the point that  the main  source of  charge quantization
phenomena,  in   the  approach   developed  in  \cite{as82},   is  the
periodicity, of the relevant variable,  due to the ring topology. With
the  increase in  the effect  of dissipation,  the particle  begins to
forget  its   ring  topology  (full  traversal  of   the  ring  become
increasingly   unlikely),   leading  to   a   suppression  of   charge
quantization phenomena.   This behavior is  similar to that  seen here
for the case of quantum walk on a cycle, where with an increase in the
effect of the environment, i.e.,  with increasing noise, the walker in
unable to perceive  the cyclic structure of the  walk space.  That the
topology-noise  interplay studied  here has  an impact  on  a concrete
condensed  matter system,  viz.   the crossover  from  strong to  weak
charge  quantization  in a  dissipative  quantum  dot, highlights  the
generality and scope of these ideas.

\section{Conclusion}
\label{conc}

We conclude that  the symmetry-topology-noise interplay presented here
would be  of relevance to quantum information  processing systems, and
have  wider implications  to the  implementation of  quantum  walks to
condensed matter systems.

\bc  {\bf  Acknowledgements} \ec  We  thank  David  Meyer for  helpful
comments.   PR  would like  to  thank  Manoj  Toshniwal for  financial
support.  CMC is thankful to  Mike and Ophelia Lazaridis for financial
support at IQC.

\end{document}